%% file: main.tex
\documentclass[final,journal,a4paper,twocolumn]{IEEEtran}

\usepackage{hyperref}   
\hypersetup{colorlinks=false,linkbordercolor=.75 .9 .95,citebordercolor=.75 .9 .95,urlbordercolor=.5 .7 .1}
\usepackage{cite}
\usepackage{amsmath,amssymb,amsfonts}
\usepackage{algorithmic}
\usepackage{graphicx}
\usepackage{textcomp}
\usepackage{xcolor}
\usepackage{enumitem}
\usepackage{subfigure} 
\usepackage{multirow}
\RequirePackage{bm}
\RequirePackage{mathrsfs}
\usepackage{booktabs} 
\usepackage{threeparttable}
\usepackage{diagbox} 
\graphicspath{{figures/}}
\usepackage{CJKutf8}

\makeatletter
\renewcommand{\maketag@@@}[1]{\hbox{\m@th\normalsize\normalfont#1}}
\makeatother

\usepackage{float}
\usepackage{algorithm}
\floatstyle{plain} 
\newfloat{myairfloat}{htbp}{myair}
{	\begin{myairfloat}[#1]
		\centering
		\begin{minipage}{#2}
			\begin{algorithm}[H]}
			{\end{algorithm}
		\end{minipage}
	\vspace{-5mm}
	\end{myairfloat}
}
\allowdisplaybreaks[4]
\newlength{\Oldarrayrulewidth}

\def\BibTeX{{\rm B\kern-.05em{\sc i\kern-.025em b}\kern-.08em
    T\kern-.1667em\lower.7ex\hbox{E}\kern-.125emX}}

\newtheorem{theorem}{\bf Proposition}

\begin{document}
\bstctlcite{myrefs:BSTcontrol}
\title{RSSI Positioning with Fluid Antenna Systems}
\author{Wenzhi Liu\textsuperscript{*}, Zhisheng Rong\textsuperscript{*}, Xiayue Liu, \\
Yufei Jiang,~\IEEEmembership{Member,~IEEE,} 
Xu Zhu,~\IEEEmembership{Senior Member,~IEEE,} 

   \thanks{*These authors contributed equally to this work.}
   
}

\maketitle

\input{body/C1}

\input{body/C2}

\input{body/C3}

        \input{body/C4}

        \input{body/C5}


\input{main.bbl}

\end{document}

%% file: body/C1.tex
\begin{abstract}
We introduce a novel received signal strength intensity (RSSI)-based positioning method using fluid antenna systems (FAS), leveraging their inherent channel correlation properties to improve location accuracy. By enabling a single antenna to sample multiple spatial positions, FAS exhibits high correlation between its ports. We integrate this high inter-port correlation with a logarithmic path loss model to mitigate the impact of fast fading on RSSI signals, and derive a simplified multipoint positioning model based on the established relationship between channel correlation and RSSI signal correlation. A maximum likelihood estimator (MLE) is then developed, for which we provide a closed-form solution. Results demonstrate that our approach outperforms both traditional least squares (LS) methods and single-antenna systems, achieving accuracy comparable to conventional multi-antenna positioning. Furthermore, we analyze the impact of different antenna structures on positioning performance, offering practical guidance for FAS antenna design.

\end{abstract}
\begin{IEEEkeywords}
	fluid antenna systems, received signal strength intensity positioning, channel correlation, maximum likelihood estimation, multi-point positioning
\end{IEEEkeywords}

\vspace{-3mm}
\section{Introduction}

The 6G mobile communication system enables emerging applications across healthcare, manufacturing, and transportation sectors\cite{yang20196g}. Key applications such as digital twins and immersive mixed reality require unprecedented positioning capabilities that include both location and attitude information. While traditional positioning technologies provide basic coordinates, they demand either extensive deployment or substantial computing resources to achieve high accuracy, resulting in significant costs.  
Consequently, Fluid Antenna Systems (FAS), a new generation of reconfigurable technology specifically designed for 6G, have gradually gained attention. FAS dynamically adjusts equivalent physical position, radiation pattern, gain, and other parameters by controlling fluid, conductive, or dielectric structures through software \cite{new2024tutorial}, offering higher degrees of freedom in communication without requiring special deployment, thus better accommodating diverse application scenarios.

Early FAS used movable materials like programmable droplets\cite{malinowski2020advances}, but their slow switching speeds limited positioning capabilities. Newer pixel-based designs use electronic switches with microsecond configuration speeds\cite{zhang2024pixel}, making embedded FAS deployment practical. Our research builds on these next-generation technologies.

The characteristics of FAS demonstrate immense potential in sensing and positioning applications. However, the specialized port design results in performance analysis that differs from conventional antennas. The correlation distribution between different ports enables fine-grained signal capture while introducing novel challenges for model analysis, becoming a key focus of current research. 

To complete theoretical analysis in sensing and related fields, various scholars have conducted a series of studies on channel models. The Jakes model describes spatial correlation between different ports \cite{chen2020massive}, but its high complexity creates significant difficulties for subsequent calculations.  To address this, many studies have approximated this model to obtain simpler representations. One such derivative, the average correlation model, eliminates the impact of the reference port and explains the relationship between the performance and size of fluid antennas\cite{wong2022closed}. The introduction of the block space concept has made the theory more applicable to cases where N approaches infinity \cite{ramirez2024new}. Meanwhile, related research has begun extending from 2D to 3D space, applicable to most application scenarios.

Current FAS channel estimation research primarily focuses on Channel State Information (CSI). CSI can  characterize frequency-selective channels more precisely while accurately capturing distance and phase information. The strong correlation between FAS ports provides robust support for recovering CSI information, demonstrating superior performance in both rich scattering environments \cite{belgiovine2021deep} and limited scattering scenarios \cite{xu2023channel}.

However, CSI imposes higher requirements on environments and transmitting devices, implying increased costs and computational demands. In comparison, while RSSI positioning has certain accuracy limitations, it shows great  potential in terms of cost, widespread applicability, and real-time implementation. The application of FAS to RSSI positioning with the aim of utilizing the continuity of signals between ports to filter out fast fading noise, enabling effective positioning in low signal-to-noise ratio environments. Additionally, we aim to reduce positioning costs and expand the range of antenna applications.

The main contributions of this paper are summarized as follows:

\begin{itemize}
   
   \item To the best of our knowledge, this is the first work to investigate RSSI positioning with FAS. Based on the channel correlation characteristics of FAS, we discuss the influence of channel correlation in RSSI positioning, and on such basis, we establish an equivalent multipoint positioning model to simplify the calculation process of FAS positioning results, and prove the effectiveness of the model.
   
    \item According to the equivalent model, we derive the maximum likelihood function of FAS positioning, and obtain the closed solution of the maximum likelihood function. Compared with the results of FAS in the least square method, single antenna and conventional multi-point positioning, it is confirmed that the measurement accuracy of MLE method can reach the result of approximate multi-point positioning.
  
    \item We analyzed the influencing factors of antenna positioning, qualitatively analyzed the impact of antenna structure on positioning effectiveness, and verified the conclusions through simulation. We found that the superiority of positioning effectiveness is positively correlated with the number of ports $N$, while the influence of the total antenna length factor $W$ on positioning effectiveness is not monotonic, and the phenomenon becomes more apparent as the antenna spacing decreases.
\end{itemize}

%% file: body/C2.tex
\section{System Model}

\begin{figure}[t]
    \centering
    \includegraphics[width=1\linewidth]{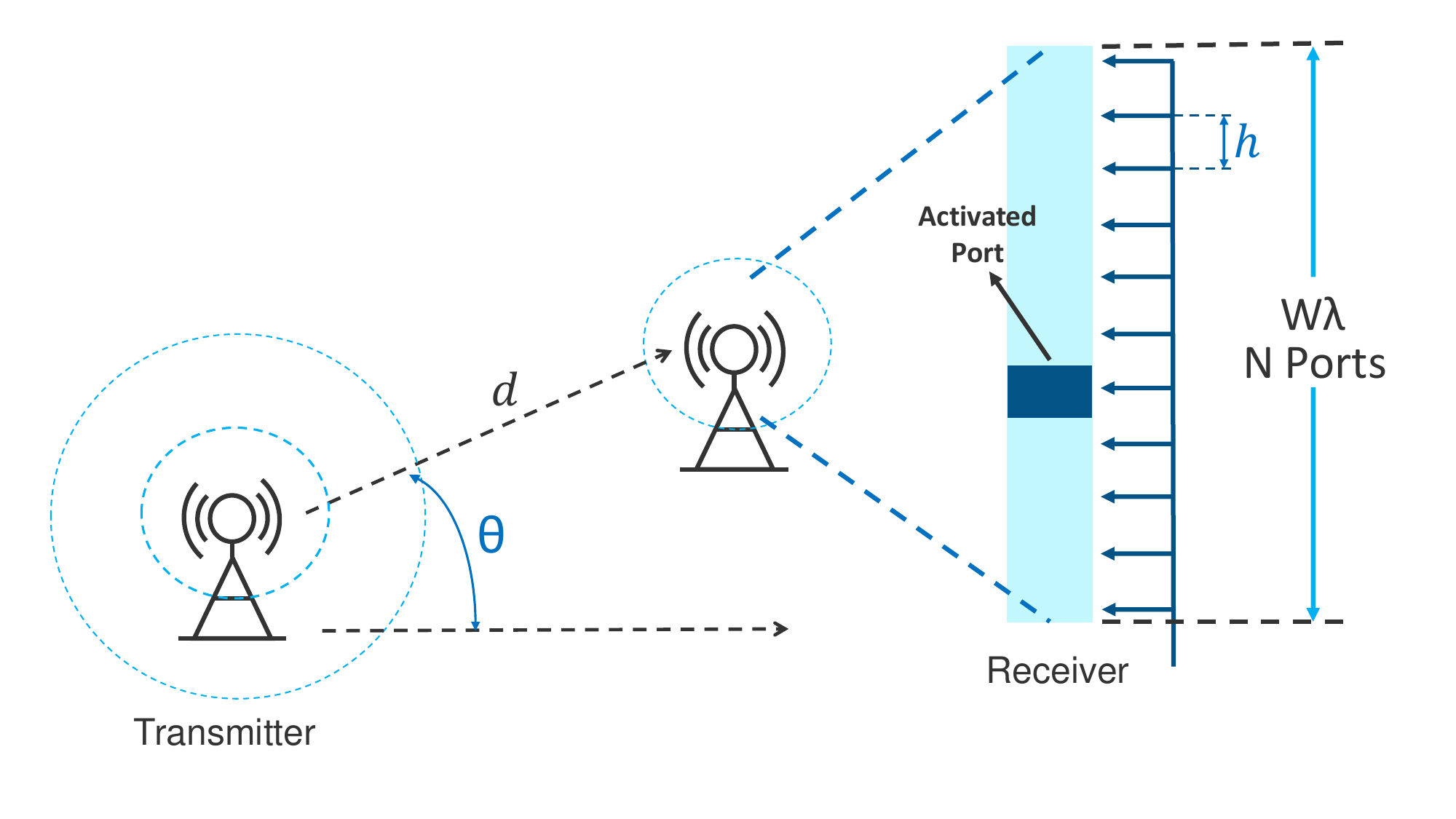}
    \caption{System model of FAS. In our positioning system model, FAS is used as the receiver and a single omnidirectional antenna is used as the transmitter. In the following, the FAS positioning result is simplified as a multi-point positioning result, and each positioning point corresponds to the port position one by one.}
    \label{fig:model2}
\end{figure}

FAS can be modeled as a single radiating element, which passes through a linear path to access the positions of $N$ different ports. The ports are uniformly distributed in the linear length, and the total arrangement length is $W \lambda $, where $W$ is the normalized length factor with respect to $\lambda $ and $\lambda $ is the signal wavelength. with the constraint that only one port is active at any given time. Therefore, FAS can be considered as a single antenna moving in a specific position on a straight line. As shown in Figure \ref{fig:model2}.

In our research, we choose a single omnidirectional antenna as the transmitter and FAS as the receiver, and assume that each port is omnidirectional. In a set of measurements, N ports are used at the same time, and the data of N ports are analyzed. We choose the latest reconfigurable antenna as the research object, and think that n groups of data are collected at approximately the same time, and treat it as a multi-point positioning method with special channel relationship. We will demonstrate the feasibility of this approximation later.

Due to $\frac{W}{N} \ll 1$, the distance $\frac{W}{N} \lambda $ between adjacent ports of FAS is small enough, and there is a strong correlation between channels, which will cause additional small-scale fading in RSSI positioning process. Next, we analyze the large-scale fading and small-scale fading models in the positioning process.

Based on the Friis transmission equation, we can calculate the received power as:
\begin{equation} \label{PR}
	P_{\textrm{R}} = P_{\textrm{T}} \frac{G_{\textrm{T}} G_{\textrm{R}} \lambda^2}{(4\pi)^2 d^n},
\end{equation}
where $P_{\textrm{T}}$ denotes the transmission power, while $G_{\textrm{T}}$ and $G_{\textrm{R}}$ represent the antenna gains of the transmitter and receiver, respectively. For our initial analysis, we consider omnidirectional antennas where the transmitting and receiving powers remain constant regardless of angle.
The distance between transmitter and receiver is denoted by $d$, and $\lambda$ denotes the signal wavelength. The path loss exponent $n$ characterizes the communication environment, accounting for shadow fading effects caused by non-line-of-sight (NLOS) propagation. This value typically ranges from 2 to 6. $n$ is not the factor that affects the final theoretical derivation result, so the path loss index in this paper is taken as 2, which can be changed according to the actual positioning environment.

For typical positioning systems, we adopt the logarithmic fading model. When incorporating Gaussian white noise, the RSSI signal strength between a transmitter and receiver separated by distance $d$ can be expressed as:
\begin{equation}\label{eq:R}
 \mathcal{R}|_{(d)} = 10 \log_{10} \left(\frac{A^2}{d^n}\right) + 30 + X_\sigma  
\end{equation}
where $A = \frac{\sqrt{P_{r}\! \left(\lambda \right)^{2} G_{T} G_{R}}}{4 \pi}$, and $X_\sigma$ represents the multipath fading effects, modeled as a zero-mean Gaussian random variable with variance $\sigma^2$, \textit{i.e.},  $X_\sigma \sim \mathcal{N}(0,\sigma^2)$.

Next, the influence of small-scale fading caused by port distribution is analyzed, and we start with the correlation between channels. Through the generalized correlation model proposed in ref \cite{beaulieu2011novel}, the fading channel parameterization $g_k$ of the port at k can be expressed as:

\begin{equation}
g_k = \sigma \left( \sqrt{1 - \mu_k^2} x_k + \mu_k x_0 \right) + j\sigma \left( \sqrt{1 - \mu_k^2} y_k + \mu_k y_0 \right), 
\end{equation}
where $x_0,x_1,...,x_N,y_0,y_1,...,y_N$ are independent and identically distributed Gaussian random variables with an average mean of zero. These parameters are used to model the relevant structure of the port channel. To simplify the mathematical analysis, the first port is chosen as the reference port.

Assuming that the receiver has two-dimensional isotropic scattering, there are differences in the phase of the arrival path between FAS ports due to spatial separation. The channel correlation distribution matrix is defined as $\mathbf{U}_{\text{g}}$ and the correlation between channels follows Jake model\cite{khammassi2023new}, as: 

 \begin{equation}\label{sum}
\begin{aligned}
 \left(\mathbf{U}_{\text{g}} \right) _{k,l} &= \text{Cov}[g_k, g_l] 
&= \sigma^2 J_0 \left( \frac{2\pi |k - l|}{(N-1)}W \right), 
\end{aligned}
\end{equation}

In the research, the first port is generally selected as the reference port. Because the ports are evenly distributed, the distance between the $k$-th port and the reference ports can be expressed as $\frac{kW}{N-1}$. The correlation $\mu_k$ between the specified port and the reference port can be expressed as
 \begin{equation}\label{mu}
  \mu_k=J_0\left(\frac{2\pi kW}{N-1}\right)
\end{equation}
 where $W$ is the normalized length of the fluid antenna relative to $\lambda$, $k=0,1,...,N-1$ and $J_0\left( \cdot \right)$ represents the first kind of zero-order Bessel function.
The weakness of the representation method is that it relies heavily on the reference port, failing to capture correlations between arbitrary port pairs. However, in the actual FAS system, if both ports are independent of the reference port, it does not mean that they must be unrelated, which introduces inaccuracies into system performance analysis. In ref \cite{wong2022closed}, a solution is proposed to optimize the spatial correlation coefficient by using the correlation coefficient $\rho$ between any two ports. The correlation coefficient $\rho_{k,l}$ between any two ports, $k$ and $l$ ($k\ne l$) can be set as follows from model (\ref{mu}):

\begin{equation}\label{rho}
  \rho_{k,l}=\mu_k\mu_l=J_0\left(\frac{2\pi \left( k-l \right)W}{N-1}\right)
\end{equation}

In order to simplify the calculation, we especially choose the average correlation coefficient model mentioned in ref \cite{wong2022closed}, and use an average parameter $\mu$ to express their correlation. This model is proved to be accurate when n is not very large\cite{ramirez2024new}, and $\mu$ can be expressed as:
\begin{equation}\label{mu:2}
  \mu ^2=\left | \frac{2}{N(N-1)} \sum_{k=1 }^{N-1} (N-k)J_0\left ( \frac{2\pi kW}{N-1}  \right )  \right | 
\end{equation}

%% file: body/C3.tex
\section{Equivalent model}

This chapter examines the ranging performance of FAS with RSSI positioning, utilizing measurement data from all ports for each measurement. We introduce a multi-point location model that simplifies FAS location result processing, shows how the model works effectively, and outlines where it can be applied. Building on this foundation, the research provides a closed-form MLE solution for FAS and explains major differences from standard multi-point measurement techniques.

A two-dimensional rectangular coordinate system is established with the FAS at the origin. The ports of the FAS are modeled as $N$ omnidirectional antennas placed at different physical locations. The position of the I-th port corresponding to the antenna is $\frac{iW}{N} \lambda$, and the distance from the port to the receiver is $d_i$. Based on the geometric relationship, the distance from the receiver to the origin $d$, is related to $d_i$ by the following equation: 
\begin{equation}\label{di}
  d_{i}^2=(\frac{iW}{N} \lambda )^2 + d^2-2\frac{iW}{N}\lambda d  \cos \theta
\end{equation}

The channel correlation between different ports can be expressed as obtained from the conclusion in the previous subsection:
\begin{equation}\label{sum:ang}
  \mathbf{R}_{\text{g}} =\begin{pmatrix}
  1&   \mu ^2&  ...&   \mu ^2 \\
  \mu ^2&  1& ... & \mu ^2 \\
  \vdots&   \vdots & \ddots &     \vdots&\\
  \mu ^2&  \mu ^2&  ...&  1&
\end{pmatrix}
\end{equation}

In actual measurement, we make FAS switch ports in turn until all N ports get data, and regard the results of N times as a group, so when processing data, we can make the following approximation:
\begin{theorem}

    When the channel correlation matrix $\mathbf{R}_{\text{g}}$ of FAS port is known, if the switching speed of FAS port is fast enough during the measurement, the measurement result can be equivalent to the measurement result of multi-point positioning model. In the multi-point positioning model, the position of the receiver corresponds to that of the FAS port one by one, and the small-scale fading generated by the port can be replaced by the correlation between signal and noise, and:
    \begin{equation} 
	\mathbf{C}_{\text{g}} =\mathbf{R}_{\text{g}}
    \end{equation}
    where $\mathbf{C}_{\text{g}}$ is the correlation matrix of RSSI signal.
\end{theorem}

\begin{IEEEproof}     
     Assuming the RSSI measurements of N ports are $ r_i $, 
     the signal model can be represented as: 
     \begin{equation} 
	{r}_{i}=P_t-20\log_{10}{d_i} +X_i
     \end{equation}
     P is the correlation matrix where RSSI is not normalized.
     $X_i$ represents the shadow fading of the i-th port, which is a Gaussian random variable with a mean of zero in this proof.
     Both the channel correlation matrix $\mathbf{R}_{\text{g}}$ and the RSSI signal correlation matrix $\mathbf{C}_{\text{g}}$ are $N\times N$ matrices, $\mathbf{C}$ is the correlation matrix where RSSI is not normalized. Since the multi-antenna system samples at the same time, the noise characteristics remain unchanged,thus $ \mathbf{C}=\sigma ^{2} \mathbf{C}_{\text{g}} $, $\sigma ^{2}$ is the noise variance at each positioning time. 
     ${(\mathbf{R}_{\text{g}})}_{i,j}$ and ${(\mathbf{C}_{\text{g}})}_{i,j}$ represents the channel correlation and RSSI signal correlation between the i-th and j-th ports, respectively. Among them:
    \begin{equation}    
	{(\mathbf{R}_{\text{g}})}_{i,j}=\frac{\text{Cov}(X_i,X_j)}{\sigma_i,\sigma_j} \end{equation}
    \begin{equation}    
        {(\mathbf{C})}_{i,j}=\text{Cov}(P_t-20\log_{10}{d_i}+X_i, P_t-20\log_{10}{d_j}+X_j)
     \end{equation}
     When the port distances are sufficiently close, like $\frac{W}{N} \ll 1$  small quantities can be neglected, thus we have:
     \begin{equation}\label{Li:ad}
     \begin{aligned}
      C_{i,j}&\approx \text{Cov}(P_t-20\log_{10}{d}+X_i, P_t-20\log_{10}{d}+X_j) \\
      & =\text{Cov}(X_i,X_j)
     \end{aligned}
     \end{equation}
     Therefore,
     $ {(\mathbf{C})}_{i,j}=\sigma _i\sigma _j{(\mathbf{R}_{\text{g}})}_{i,j}$
     and $\mathbf{C}$ can be expressed as:
     \begin{equation}    
       \mathbf{C}=\mathbf{R}_{\text{g}}\cdot \text{diag}({\sigma_1} ^{2} ,{\sigma_2} ^{2},...,{\sigma_N} ^{2})
     \end{equation}
     where $\text{diag}\left( \cdot \right)$ stands for diagonal matrix
     When all the ${\sigma_i} ^{2}$ are equal, it is obvious that the equation $\mathbf{C}_{\text{g}} =\mathbf{R}_{\text{g}}$ holds. 
     The requirement for this condition to hold is that the switching speed is fast enough so that the RSSI measurements from all ports can be considered to be obtained almost simultaneously. The port switching speed of the new generation of fluid antennas can achieve this\cite{zhang2024pixel}. It is also required that all ports experience almost the same environmental conditions, with their shadow fading exhibiting similar statistical characteristics, and condition $\frac{W}{N} \ll 1$ ensures this. Under these assumptions, the correlation between port channels can be approximately determined by the correlation between the RSSI signals received from each port.
\end{IEEEproof}

\begin{theorem}
   In the equivalent multi-point positioning model, $d$ can be solved by the following formula, which is in the form of weighted sum:
    \begin{equation} \label{biMi}   
	\sum_{i=1}^{N}b[i]x[i]= \sum_{i=1}^{N}b[i]M_i(d,\theta ) 
    \end{equation}
    where $M_{i} (d,\theta )=30-20\log_{10} (\frac{d_i}{A} ) $, $x[i]=M_{i}(d,\theta ) +X_{\sigma }$, and $X_{\sigma } $ represents white Gaussian noise in the channel.
   In addition, the change of weight $b[i] $ is affected by the correlation between signals and can be expressed as:
   \begin{equation}\label{bi}
     b[i]=\frac{\partial M_i(d,\theta )}{\partial d} -k\sum_{i=1}^{N} \frac{\partial M_i(d,\theta )}{\partial d} 
   \end{equation}
   When there is no correlation, this result can be reduced to MLE expression of traditional antenna array. 
\end{theorem} 
\begin{IEEEproof}
   According to the characteristics of Gaussian white noise, $x[i]\sim (M_i(d,\theta ),\sigma ^2)$ and the likelihood function $L(d,\theta)$ can be given by
   \begin{equation}\label{Li}
      L(d,\theta )=\frac{1}{(2\pi )^\frac{N}{2}\left | \mathbf{C}_{\text{g}}  \right |^\frac{1}{2}} \exp\left [ -\frac{1}{2} (\bar{P}-P_{0}  )^{T} \mathbf{C}_{\text{g}} ^{-1} (\bar{P}-P_{0}) \right ] 
   \end{equation}

  According to the above proof, $\mathbf{C}_{\text{g}}$ is equal to $\mathbf{R}_{\text{g}}$.

  $\bar{P}$ is the $N\times 1$ order vector matrix corresponding to the actually received signal $x[i]$ of $N$ ports, and $P_{0}$ is the $N\times 1$ order vector matrix corresponding to $M_i(d,\theta)$.

Let $a=\mu^2$, substituting the above results into (\ref{Li}) can be obtained:
\begin{equation}\label{Li:ad}
\begin{aligned}
&\small{L(d,\theta)}= \small{\frac{1}{(2\pi)^\frac{N}{2} (1-a)^\frac{N-1}{2} \left[1 + (N-1)a\right]^\frac{1}{2}}} \\
&\quad \small{\times \exp\left\{ -\frac{1}{2} \left[ \frac{1}{1-a^2} \sum_{i=1}^{N} x_i^2 - \frac{1}{(1-a^2)} k \right] \left( \sum_{i=1}^{N} x_i \right)^2 \right\}}
\end{aligned}
\end{equation}
where $k=\frac{a^2}{(1-a^2)\left [ 1+a^2(N-1) \right ] }$, and $x_i$ is Gaussian noise on different port channels.

The logarithm of the above formula can be obtained:
\begin{equation}\label{Li:ln}
\begin{aligned}
\small&{\ln L(d,\theta)} = \small{-\frac{1}{2} \ln \left\{ (2\pi)^N (1-a)^{N-1} \left[ 1 + (N-1)a \right] \right\}} \\
&\quad \small{-\frac{1}{2} \left\{ \frac{1}{1-a^2} \sum_{i=1}^{N} x_i^2 - \frac{1}{1-a^2} k \left( \sum_{i=1}^{N} x_i \right)^2 \right\}}
\end{aligned}
\end{equation}

Let
\begin{equation}\label{E}
\begin{aligned}
&{E(d,\theta )=\frac{1}{1-a^2} \sum_{i=1}^{N} x_i^2 -\frac{1}{1-a^2}k( \sum_{i=1}^{N} x_i)^2}
\end{aligned}
\end{equation}

The extreme value of $E(d,\theta )$ should appear at the same point as the maximum value of likelihood function, which means:
\begin{equation}\label{E}
  \frac{\partial E(d,\theta )}{\partial d} =0 , \frac{\partial E(d,\theta )}{\partial \theta} =0
\end{equation}

In this paper, only the parameter $d$ is discussed. Substituting $E(d,\theta )$ into the partial derivative of d can be obtained:
\begin{equation}
\begin{aligned}\label{M}
&\sum_{i=1}^{N} \left \{ [x[i] - M(d,\theta)] \frac{\partial M(d,\theta)}{\partial d} \right \} \\&= k \left \{ \sum_{i=1}^{N} \left [ x[i] - M(d,\theta) \right ] \right \} 
 \times \sum_{i=1}^{N} \frac{\partial M(d,\theta)}{\partial d}
\end{aligned}
\end{equation}

In the above formula:
\begin{equation}\label{E}
 \frac{\partial M_i(d,\theta )}{\partial d} = -\frac{10}{\ln_{}{10} } \frac{2d-\frac{2iW}{N}\lambda \cos \theta  }{d^2-\frac{2iW}{N}\lambda d\cos \theta  } 
\end{equation}

High-order small quantity has been ignored.

 If let 
 $$b[i]=\frac{\partial M_i(d,\theta )}{\partial d} -k\sum_{i=1}^{N} \frac{\partial M_i(d,\theta )}{\partial d} $$

Obviously, $B=\left \{ b[1],b[2],...,b[N] \right \} $ is an array that is independent of $d$ and $\theta$.
The above results can be simplified to the form of weighted sum as (\ref{biMi}).

\end{IEEEproof}

%% file: body/C4.tex
\section{Simulation results} \label{book:res}

\begin{figure}[t]
    \centering
    \includegraphics[width=0.8\linewidth]{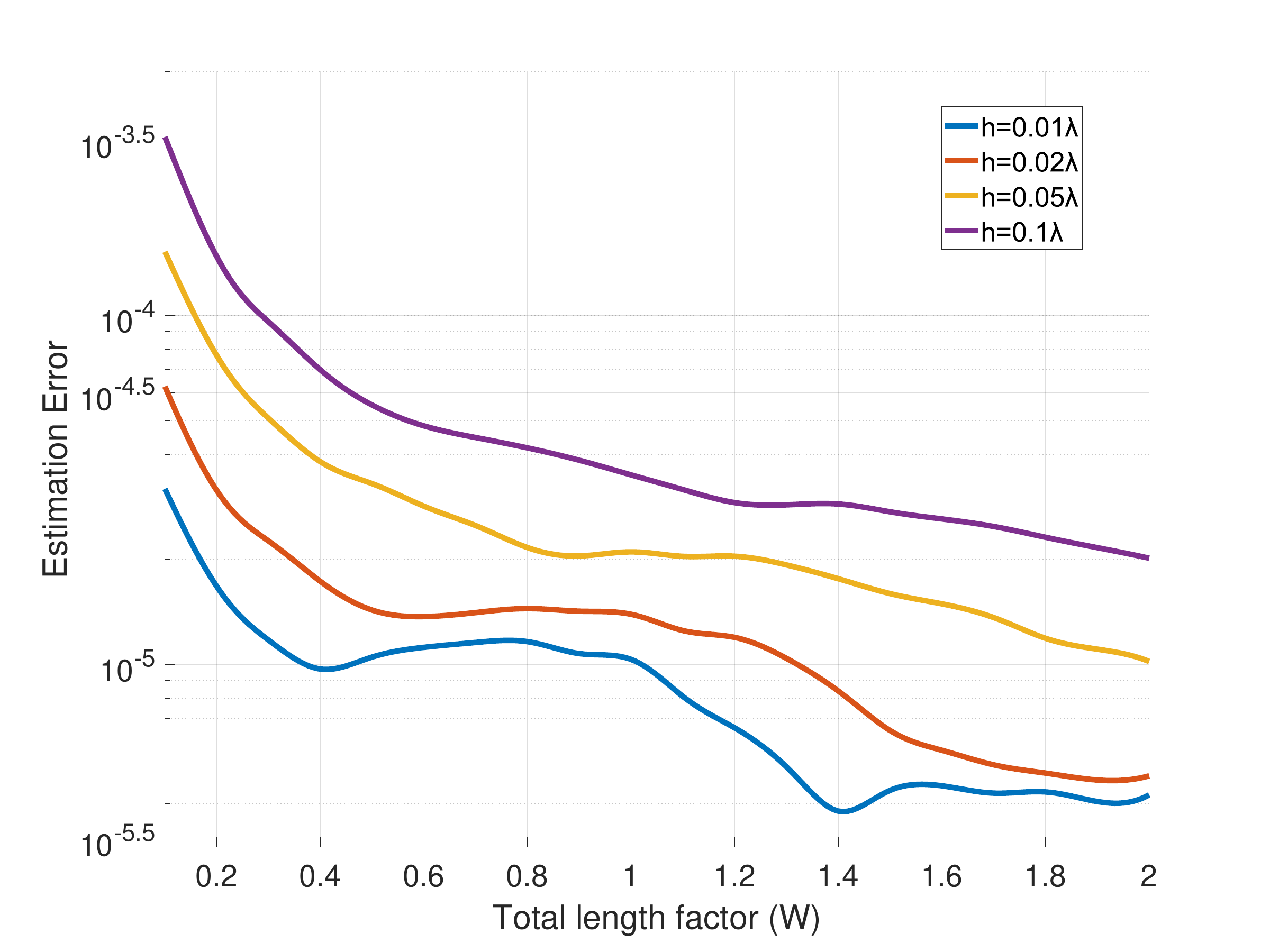}
    \caption{Schematic diagram of the influence of antenna design parameters on measurement results. $SNR=10dB$, $h=\frac{W}{N}\lambda $, LS as the calculation method and NMSE as the error definition method.}
    \label{fig:WN}
\end{figure}

In this section, we compare the positioning results of MLE method and LS method of FAS, single omnidirectional antenna and LS method of multi-point positioning, and make a qualitative analysis of the comparison results in \ref{subsection:BBC}. At the same time, we analyze the influence of antenna structure on the positioning effect in \ref{subsection:CCC}. The final results are shown in Figs. \ref{fig:MLE_LS} and \ref{fig:WN}.

\subsection{Performance comparison} \label{subsection:BBC}

In this chapter, different antenna types are compared with different calculation methods. In order to unify the variables, the number of FAS ports is the same as that of antennas in multipoint positioning, and the number of measurements of a single antenna is set to $N$ times that of FAS and multipoint positioning. 

\begin{figure}[t]
    \centering
    \includegraphics[width=0.8\linewidth]{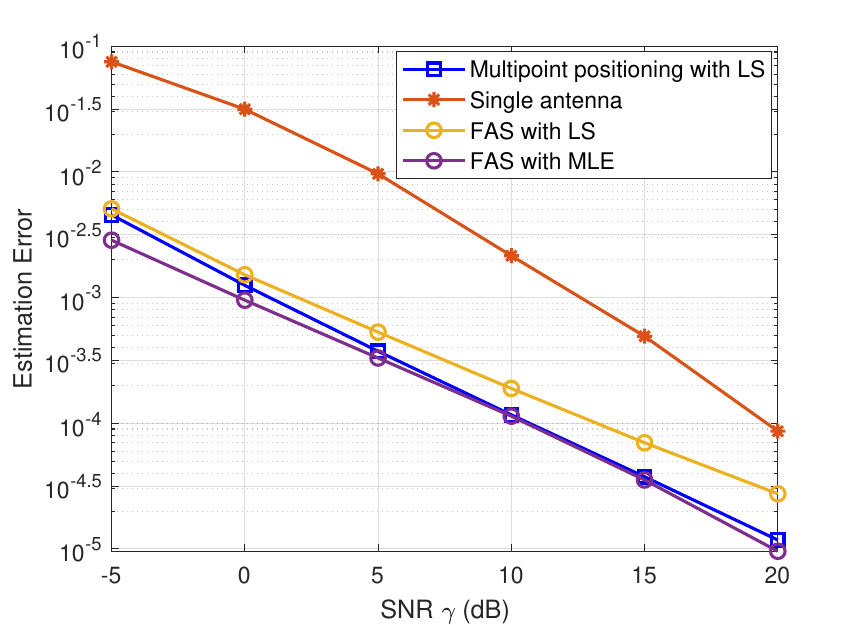}
    \caption{Comparing the results of FAS, multipoint positioning and single antenna RSSI positioning, where $N = 12, W = 0.5$, LS stands for least square method, MLE stands for maximum likelihood function, and the error is defined as NMSE.}
    \label{fig:MLE_LS}
\end{figure}

As can be seen from Figure \ref{fig:MLE_LS}, the measurement accuracy of FAS is more than 10dB higher than that of ordinary single antenna. Although the positioning performance of FAS under LS algorithm is slightly lower than that of traditional multi-point positioning, it can approach the result of multi-point positioning well under MLE method, and the measurement result of MLE method is better than that of LS method. The results show that FAS has great potential in positioning compared with ordinary single antenna, and can achieve better performance at lower cost. The performance of FAS in small signal-to-noise ratio shows that the continuous performance of signals between ports can effectively weaken the influence of small-scale fading on the original signal. However, because the ports are arranged in a straight line and the signals are not independent of each other, FAS has inherent defects in obtaining the richness of positioning signals, so when LS is also selected, there is still a gap of about 2dB between FAS and multi-point positioning.

\subsection{Impact factor analysis} \label{subsection:CCC}

This section discusses the influencing factors of fluid antenna positioning results, makes qualitative analysis, and preliminarily analyzes the best parameters to reduce errors. Where $h$ corresponds to the interval between adjacent antennas.

As can be seen from the Figure \ref{fig:WN}, under the condition of constant $ w $,with the doubling of $ n $,the positioning accuracy is steadily improved by about 2.5dB at one time. This is because with the increase of $N$, the antenna spacing decreases and the correlation between ports increases, which can capture the changes of fast fading signals more completely and reduce noise interference. It is worth mentioning that when $W$ increases and $h$ decreases to a certain extent, the average correlation coefficient model cannot fully meet the actual demand because of the large number of $N$.

The influence of length factor $W$ on the measurement results is not monotonous. When the distance between adjacent ports is fixed, with the increase of the total antenna length, although the positioning error generally decreases with the increase of $W$, it is not monotonous in a small area, and its fluctuation is closely related to the correlation, which is close to the form of Bessel function. This fluctuation gradually weakens with the increase of $N$ and antenna spacing. When $h=0.05\lambda$, the overall change trend is nearly monotonous, which is the case with common multi-point positioning. However, when $h$ tends to be below $0.01\lambda$, this change becomes obvious, and an obvious extreme value appears at $W=0.5$. At present, the design concept of FAS pursues the situation that $N$ tends to infinity, so such small-scale fluctuation provides inspiration for antenna design.

%% file: body/C5.tex
\section{Conclusion} \label{book:con}
In our paper, the application of FAS in RSSI location is studied, and its advantages over ordinary single antenna are revealed. Starting with the average correlation coefficient model of the channel, we propose a multi-point positioning model to analyze the measurement results of FAS to simplify the calculation results, and give the closed-form solution of MLE.
The analysis proves that the correlation between ports based on FAS effectively reduces the influence of small-scale fading signals in RSSI measurement process, so that its measurement accuracy can well approach the measurement accuracy of traditional multipoint positioning, and MLE shows greater advantages than LS. It is possible to create a virtual antenna array without complicated hardware.
Our research on the influence of antenna structure shows that the positioning accuracy increases with the increase of port number $N$, while the total antenna length $W$ has a non-monotonic influence, and a significant extreme value is observed when $W=0.5$, especially at a small interval related to the future development of FAS. These findings provide valuable design guidelines.
It is worth mentioning that, although this method is limited to N cases without considering the block space, it still confirms the potential of FAS in RSSI positioning, which can achieve accurate positioning in resource-limited applications and provide a good foundation for the Internet of Things, robotics and mobile communication.

%% file: main.bbl

%% file: main.bbl
\begin{thebibliography}{10}
\providecommand{\url}[1]{#1}
\csname url@samestyle\endcsname
\providecommand{\newblock}{\relax}
\providecommand{\bibinfo}[2]{#2}
\providecommand{\BIBentrySTDinterwordspacing}{\spaceskip=0pt\relax}
\providecommand{\BIBentryALTinterwordstretchfactor}{4}
\providecommand{\BIBentryALTinterwordspacing}{\spaceskip=\fontdimen2\font plus
\BIBentryALTinterwordstretchfactor\fontdimen3\font minus
  \fontdimen4\font\relax}
\providecommand{\BIBforeignlanguage}[2]{{%
\expandafter\ifx\csname l@#1\endcsname\relax
\typeout{** WARNING: IEEEtran.bst: No hyphenation pattern has been}%
\typeout{** loaded for the language `#1'. Using the pattern for}%
\typeout{** the default language instead.}%
\else
\language=\csname l@#1\endcsname
\fi
#2}}
\providecommand{\BIBdecl}{\relax}
\BIBdecl
\renewcommand{\BIBentryALTinterwordstretchfactor}{4}

\bibitem{yang20196g}
P.~Yang, Y.~Xiao, M.~Xiao \emph{et~al.}, ``6g wireless communications: Vision
  and potential techniques,'' \emph{IEEE network}, vol.~33, no.~4, pp. 70--75,
  Nov. 2019.

\bibitem{new2024tutorial}
W.~K. New, K.-K. Wong, H.~Xu \emph{et~al.}, ``A tutorial on fluid antenna
  system for 6g networks: Encompassing communication theory, optimization
  methods and hardware designs,'' \emph{IEEE Communications Surveys \&
  Tutorials}, Nov. 2024.

\bibitem{malinowski2020advances}
R.~Malinowski, I.~P. Parkin, and G.~Volpe, ``Advances towards programmable
  droplet transport on solid surfaces and its applications,'' \emph{Chemical
  Society Reviews}, vol.~49, no.~22, pp. 7879--7892, Jun. 2020.

\bibitem{zhang2024pixel}
J.~Zhang, J.~Rao, Z.~Li \emph{et~al.}, ``A novel pixel-based reconfigurable
  antenna applied in fluid antenna systems with high switching speed,''
  \emph{IEEE Open Journal of Antennas and Propagation}, vol.~6, no.~1, pp.
  212--228, Nov. 2025.

\bibitem{chen2020massive}
X.~Chen, D.~W.~K. Ng, W.~Yu \emph{et~al.}, ``Massive access for 5g and
  beyond,'' \emph{IEEE Journal on Selected Areas in Communications}, vol.~39,
  no.~3, pp. 615--637, Sep. 2020.

\bibitem{wong2022closed}
K.~Wong, K.~Tong, Y.~Chen \emph{et~al.}, ``Closed-form expressions for spatial
  correlation parameters for performance analysis of fluid antenna systems,''
  \emph{Electronics Letters}, vol.~58, no.~11, pp. 454--457, Apr. 2022.

\bibitem{ramirez2024new}
P.~Ram{\'\i}rez-Espinosa, D.~Morales-Jimenez, and K.-K. Wong, ``A new spatial
  block-correlation model for fluid antenna systems,'' \emph{IEEE Transactions
  on Wireless Communications}, Aug. 2024.

\bibitem{belgiovine2021deep}
M.~Belgiovine, K.~Sankhe, C.~Bocanegra \emph{et~al.}, ``Deep learning at the
  edge for channel estimation in beyond-5g massive mimo,'' \emph{IEEE Wireless
  Communications}, vol.~28, no.~2, pp. 19--25, May 2021.

\bibitem{xu2023channel}
H.~Xu, G.~Zhou, K.-K. Wong \emph{et~al.}, ``Channel estimation for fas-assisted
  multiuser mmwave systems,'' \emph{IEEE Communications Letters}, vol.~28,
  no.~3, pp. 632--636, Dec. 2023.

\bibitem{beaulieu2011novel}
N.~C. Beaulieu and K.~T. Hemachandra, ``Novel simple representations for
  gaussian class multivariate distributions with generalized correlation,''
  \emph{IEEE Transactions on Information Theory}, vol.~57, no.~12, pp.
  8072--8083, Dec. 2011.

\bibitem{khammassi2023new}
M.~Khammassi, A.~Kammoun, and M.-S. Alouini, ``A new analytical approximation
  of the fluid antenna system channel,'' \emph{IEEE Transactions on Wireless
  Communications}, vol.~22, no.~12, pp. 8843--8858, Apr. 2023.

\end{thebibliography}
